\documentclass[
twocolumn,
preprintnumbers,
nofootinbib,
prd,aps
]{revtex4}

\usepackage{epsf}
\usepackage{latexsym}
\usepackage{amssymb}
\usepackage{epsf}
\usepackage{amsmath}
\usepackage{slashed}

\begin{document}


\newcommand{\rem}[1]{$\spadesuit${\bf #1}$\spadesuit$}

\renewcommand{\topfraction}{0.8}

\preprint{UT-HET 102, UT-15-24}

\title{

Beam Dump Experiment at Future Electron-Positron Colliders

}

\author{
Shinya Kanemura$^{(a)}$, Takeo Moroi$^{(b)}$, Tomohiko Tanabe$^{(c)}$
}

\affiliation{
$^{(a)}$Department of Physics, University of Toyama,
Toyama 930-8555, Japan
\\
$^{(b)}$Department of Physics, The University of Tokyo, 
Tokyo 113-0033, Japan
\\
$^{(c)}$ICEPP, The University of Tokyo, 
Tokyo 113-0033, Japan
}

\date{July, 2015}

\begin{abstract}

  We propose a new beam dump experiment at future colliders with
  electron ($e^-$) and positron ($e^+$) beams, BD$ee$, which will
  provide a new possibility to search for hidden particles, like
  hidden photon.  If a particle detector is installed behind the beam
  dump, it can detect the signal of in-flight decay of the hidden
  particles produced by the scatterings of $e^\pm$ beams off materials
  for dumping.  We show that, compared to past experiments, BD$ee$ (in
  particular BD$ee$ at $e^+e^-$ linear collider) significantly
  enlarges the parameter region where the signal of the hidden
  particle can be discovered.

\end{abstract}

\maketitle

\renewcommand{\thefootnote}{\#\arabic{footnote}}

High energy colliders with electron ($e^-$) and positron ($e^+$)
beams, such as the International Linear Collider (ILC) \cite{ILC-TDR},
the Compact Linear Collider (CLIC) \cite{CLIC}, and Future Circular
Collider with $e^+e^-$ beams (FCC-$ee$) \cite{FCCee}, are widely
appreciated as prominent candidates of future experiments.  One of the
reasons is that, with the discovery of Higgs boson at the LHC
\cite{HiggsDiscovery}, detailed studies of Higgs properties at
$e^+e^-$ colliders are now very important \cite{Peskin:2013xra}.  In
addition, $e^+e^-$ colliders have sensitivity to new particles at TeV
scale or below if they have electroweak quantum numbers.

Although $e^+e^-$ colliders have many advantages in studying physics
beyond the standard model (BSM), they can hardly probe BSM particles
whose interaction is very weak.  We call such particles \emph{hidden
  particles}, which appear in various BSM models.  For example, there
may exist a gauge symmetry other than those of the standard model
(SM), as is often the case in string theory.  If the breaking scale of
such a hidden gauge symmetry is lower than the electroweak scale, the
associated gauge boson can be regarded as a hidden particle
\cite{DarkPhoton}.  In string theory, it has also been pointed out
that there may exist axion-like particles (ALPs)
\cite{Arvanitaki:2009fg}; they are also candidates of the hidden
particle.  Sterile neutrino is another example.  These particles
interact very weakly with SM particles, and are hardly accessed by
studying $e^+e^-$ collisions.  If $e^+e^-$ colliders will be built in
the future, it is desirable to make it possible to study hidden
particles as well.

In this letter, we discuss a possibility to detect hidden particles at
the $e^+e^-$ facilities.  We propose a beam dump experiment at future
$e^+e^-$ colliders (BD$ee$), in which the beam after the $e^+e^-$
collision is used for the beam dump experiment.  In particular, at the
ILC and CLIC, the $e^\pm$ beams will be dumped after each collision,
which makes a large number of $e^\pm$ available for the beam dump
experiment.  Using the hidden photon, which is the gauge boson
associated with a (spontaneously broken) hidden $U(1)$ symmetry, as an
example, we show that the BD$ee$ can cover a parameter region which
has not been explored by past experiments.

Let us first summarize the basic setup of BD$ee$.  We simply assume
the current design of the beam dump system of the ILC although one may
consider other possibilities.  The main beam dumps of the ILC will
consist of $1.8\ {\rm m}$-diameter cylindrical stainless-steel
high-pressure ($10\ {\rm bar}$) water vessels \cite{ILC-TDR}.  The
$e^\pm$ beams after passing through the interaction point are injected
into the dump, which absorbs the energy of the electromagnetic shower
in $11\ {\rm m}$ of water.  If there exists a hidden particle, like
hidden photon, for example, it is produced by the $e^\pm$-${\rm H_2O}$
scattering process.  In this letter, to make our discussion concrete,
we consider the case where the target is ${\rm H_2O}$, although other
materials may be used as a target.  The number of the hidden photon
produced in the dump is insensitive to the target material.

\begin{figure}[t]
  \centerline{\epsfxsize=0.475\textwidth\epsfbox{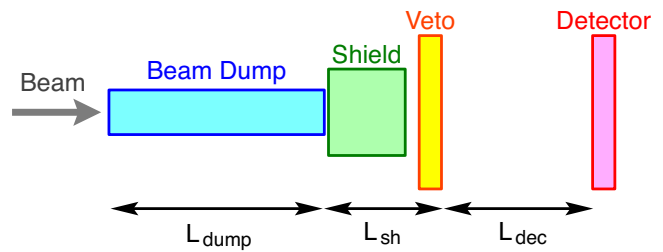}}
  \caption{Schematic view of BD$ee$. The electron (or positron) beam
    is injected into the beam dump from the left.}
  \label{fig:setup}
\end{figure}

Our proposal is to install a particle detector behind the dump, with
which we can observe signals of hidden particles produced in the dump.
The schematic picture of the setup of BD$ee$ is shown in Fig.\
\ref{fig:setup}.  The decay volume is a vacuum vessel with the length
of $L_{\rm dec}$; the signal of the hidden particle is detected if the
hidden particle decays into (visible) SM particles in the decay
volume.  A tracking detector is used to detect the hidden particle
decaying into a pair of charged particles.  Additional detectors such
as calorimeters and muon detectors may be installed to enrich the
physics case.  As well as the hidden particles, charged particles are
also produced in the dump; rejection of those particles is essential
to suppress backgrounds.  In particular, a significant amount of muons
are produced, as we will discuss in the following.  Thus, we expect to
install shields and veto counters between the dump and the decay
volume.  Additional veto counters surrounding the detector serve to
reject cosmic rays.  

To see the sensitivity of BD$ee$, we consider a model with hidden
photon (denoted as $X$), which has a small kinetic mixing with
ordinary photon.  We adopt the following Lagrangian in our analysis
\begin{align}
  {\cal L} = &
  {\cal L}_{\rm SM}
  -\frac{1}{4} F_{\mu\nu}^{(X)} F_{\mu\nu}^{(X)}
  -\frac{\epsilon}{2} F_{\mu\nu}^{\rm (em)} F_{\mu\nu}^{(X)}
  + \frac{m_X^2}{2} X_\mu X_\mu,
\end{align}
where ${\cal L}_{\rm SM}$ is the SM Lagrangian, and $F_{\mu\nu}^{\rm
  (em)}$ and $F_{\mu\nu}^{(X)}$ are field strength tensors of
electromagnetic and hidden photons, respectively.  In addition,
$\epsilon$ is the mixing parameter, which is assumed to be much
smaller than $1$, while $m_X$ is the mass of the hidden photon.

Once the hidden photon is produced in the dump, it may go through the
shield region because the hidden photon is very weakly interacting,
and may decay into SM particles in the decay volume.  Thus, the SM
particles (like a pair of charged particles) originating from the
decay volume are the signal of the hidden photon production.

The hidden photon production is dominated by the $t$-channel
(ordinary) photon exchange process of $e^\pm N\rightarrow e^\pm X N'$,
where $N$ is a nucleus in ${\rm H_2O}$ while $N'$ denotes the hadrons
in the final state.  Because of the masslessness of the photon, the
cross section is enhanced for the configuration in which
$\tilde{t}\equiv -q^2$ takes its minimal possible value, where
$q\equiv P_N-P_{N'}$ denotes the momentum of the virtual photon, with
$P_N$ and $P_{N'}$ being the momenta of $N$ and $N'$, respectively.
(The $\tilde{t}$ parameter should not be confused with the length in
units of the radiation length, which will be denoted as $t$ in this
letter.)  Consequently, the hidden photon $X$ is likely to be emitted
in (almost) the beam direction, which will be taken to be the
$z$-axis.  Because we are interested in the case where $E_e\gg m_X$,
the decay products of $X$ are also likely to be emitted in (almost)
the beam direction.  Thus, the particle detector behind the dump can
efficiently observe the decay products of $X$.

Using Weizs\"acker-Williams approximation, the cross section for
$e^\pm N\rightarrow e^\pm X N'$ is estimated as $d \sigma(e^\pm
N\rightarrow e^\pm X N')/dx=\epsilon^2 d \sigma_0/dx$, where
\cite{Kim:1973he, Bjorken:2009mm, Andreas:2012mt}
\begin{align}
  \frac{d \sigma_0}{dx}
  = 4 \alpha^3 \chi \beta_X
  \left( 1 - x + \frac{1}{3}x^2 \right)
  \left( \frac{1-x}{x} m_X^2 + x m_e^2 \right) ^{-1}. 
\end{align}
Here, $\alpha$ is the QED fine structure constant, $m_e$ is the
electron mass, $x\equiv E_X/E_e$, $\beta_X\equiv\sqrt{1 -
  (m_X^2/E_e^2)}$, and $\chi$ is the effective flux of photons.  In
our numerical analysis, we use 
\begin{align}
  \chi = \int_{\tilde{t}_{\rm min}}^{\tilde{t}_{\rm max}} d \tilde{t}
  \frac{\tilde{t} - \tilde{t}_{\rm min}}{\tilde{t}^2} 
  G_2 (\tilde{t}),
\end{align}
where $\tilde{t}_{\rm min}=(m_X^2/2E_e)^2$, and $\tilde{t}_{\rm
  max}=m_X^2$.  For a nucleus with the charge $Z$, the electric form
factor is given by \cite{Bjorken:2009mm}
\begin{align}
  G_2 (\tilde{t}) = &
  \left( \frac{a^2 \tilde{t}}{1 + a^2 \tilde{t}} \right)^2
  \left( \frac{1}{1 + \tilde{t}/d} \right)^2 Z^2  
  \nonumber \\ &
  + \left( \frac{a'^2 \tilde{t}}{1 + a'^2 \tilde{t}} \right)^2
  \left( 
    \frac{1 + (\mu_p^2-1)\tilde{t}/4m_p^2}{(1 + \tilde{t}/d')^4} 
  \right)^2 Z,
  \label{G2}
\end{align}
where $m_p$ is the proton mass, $a=111Z^{-1/3}/m_e$, $d=0.146\ {\rm
  GeV}^2 A^{-2/3}$ (with $A$ being the atomic number),
$a'=773Z^{-2/3}/m_e$, $d'=0.71\ {\rm GeV}^2$, and $\mu_p=2.79$.  (The
first and the second terms of the right-hand side of Eq.\ \eqref{G2}
represent elastic and inelastic components, respectively.)

After the injection into the dump, the beam loses its energy.  We use
the following energy distribution of $e^-$ after passing through a
medium of the radiation length $t$ \cite{Tsai:1986tx}:
\begin{align}
  I_e(E_{\rm beam},E_e,t) = \frac{1}{E_{\rm beam}}
  \frac{[\ln (E_{\rm beam}/E_e)]^{bt - 1}}{\Gamma (bt)},
\end{align}
where $E_{\rm beam}$ is the energy of the electron beam just before
the injection into the dump, and $b=\frac{4}{3}$.

The total number of the signal is given by \cite{Bjorken:2009mm,
  Andreas:2012mt}
\begin{widetext}
\begin{align}
  N_{\rm sig} = N_e \frac{N_{\rm Avo} X_0}{A} \epsilon^2 
  B_{\rm sig}
  \int_{m_X}^{E_{\rm beam}-m_e} dE_X \int_{E_X+m_e}^{E_{\rm beam}} dE_e 
  \int_0^T dt 
  \frac{I_e(E_{\rm beam},E_e,t)}{E_e}
  \left. \frac{d\sigma_0}{dx} \right|_{x=E_X/E_e}
  P_{\rm dec},
  \label{Nsig}
\end{align}
\end{widetext}
where $N_e$ is the total number of electron injected into the dump,
$N_{\rm Avo}$ is the Avogadro constant, $X_0\simeq
716.4A/[Z(Z+1)\ln(287/\sqrt{Z})]\ {\rm g/cm^2}$ is the radiation length,
$T\equiv \rho L_{\rm dump}/X_0$ with $\rho$ being the density of
water, and $B_{\rm sig}$ is the branching ratio of $X$ into the signal
channel.  (Hereafter, for simplicity, we take $B_{\rm sig}=1$.)  In
addition, $P_{\rm dec}$ is the probability of the decay of $X$ in the
decay volume.  With the present setup, $L_{\rm dump}$ is so long that
the hidden photon production mostly occurs near the edge of the dump.
Thus, we approximate
\begin{align}
  P_{\rm dec} = 
  e^{-(L_{\rm dump}+L_{\rm sh})/l_X} (1-e^{-L_{\rm dec}/l_X}),
\end{align}
where $l_X$ is the decay length of $X$ with energy $E_X$.  Using
$R\equiv\sigma(e^+e^-\rightarrow\mbox{hadrons})/
\sigma(e^+e^-\rightarrow\mu^+\mu^-)$, we evaluate $l_X^{-1}$ as
\begin{align}
  l_X^{-1} = \frac{m_X}{E_X} 
  \left[
    \sum_{\ell=e,\mu,\tau}
    \Gamma_{X\rightarrow \ell^+ \ell^-}
    + R \Gamma_{X\rightarrow \mu^+ \mu^-}
  \right].
\end{align}
(In our numerical calculation, we use $R$ given in
\cite{Agashe:2014kda}.)  The decay rate of $X$ into a lepton pair is
given by
\begin{align}
  \Gamma_{X\rightarrow \ell^+ \ell^-} = 
  \frac{\alpha \epsilon^2}{3} m_X
  \left( 1 + \frac{2m_\ell^2}{m_X^2} \right)
  \sqrt{ 1 - \frac{4m_\ell^2}{m_X^2} },
\end{align}
with $m_\ell$ being the mass of the lepton $\ell$.

The number of events is proportional to the total number of injected
electrons which depends on collider parameters.  First, we consider
the case of the ILC, at which electron and positron beams are dumped
immediately after passing thought the interaction point.  (We call
such a case ``BD$ee$LC.'')  In the current design of the ILC, the
bunch train consists of 1312 bunches, each of which contains $2\times
10^{10}$ electrons, and is dumped with the frequency of $5\ {\rm Hz}$
\cite{ILC-TDR}.  Thus, with one-year (i.e., $3\times 10^7\ {\rm sec}$)
operation, about $4\times 10^{21}$ electrons are injected into the
dump.  While we take this value as the basis for our calculation,
luminosity upgrades are foreseen in the later stage of the ILC
operation, which doubles the number of bunches \cite{Barklow:2015tja}.
In the case of CLIC, a similar calculation yields $(2-4)\times
10^{21}$ electrons using the parameters given in \cite{CLIC}.  Since
these numbers are similar in order of magnitude, in the following
discussion, we take the ILC number and scale the beam energy up to the
CLIC energy range.

\begin{figure}[t]
  \centerline{\epsfxsize=0.475\textwidth\epsfbox{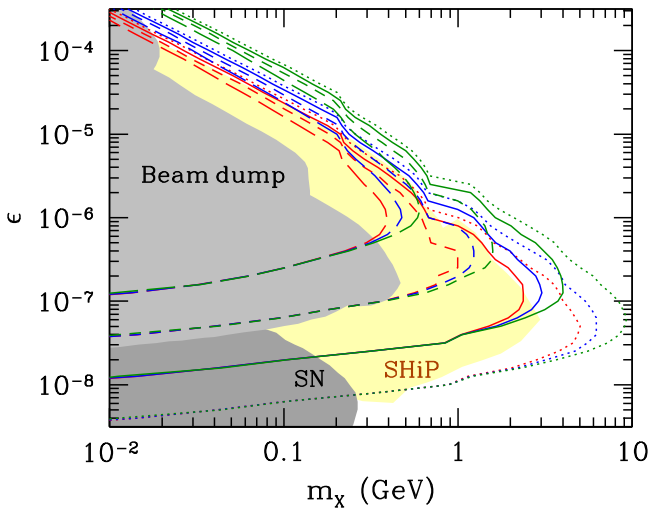}}
  \caption{Contours of constant $N_{\rm sig}$ on the $m_X$ vs.\
    $\epsilon$ plane for $E_{\rm beam}=250$ (red), $500$ (blue), and
    $1500\ {\rm GeV}$ (green), taking $N_e=4\times 10^{21}$, $L_{\rm
      dump}=11\ {\rm m}$, $L_{\rm sh}=50\ {\rm m}$, and $L_{\rm
      dec}=50\ {\rm m}$.  The dotted, solid, short-dashed, and
    long-dashed lines correspond to $N_{\rm sig}=10^{-2}$, $1$,
    $10^2$, and $10^4$, respectively.  The gray-shaded regions are
    already excluded by past beam dump experiments
    \cite{Andreas:2012mt} (light-gray) or supernova bounds
    \cite{Kazanas:2014mca} (dark-gray), while SHiP experiment, if
    approved, will cover the yellow-shaded one
    \cite{Alekhin:2015byh}.}
  \label{fig:mxvseps}
\end{figure}

We numerically integrate Eq.\ \eqref{Nsig} to evaluate the number of
events.  Taking $N_e=4\times 10^{21}$, $L_{\rm dump}=11\ {\rm m}$,
$L_{\rm sh}=50\ {\rm m}$, and $L_{\rm dec}=50\ {\rm m}$, we calculate
$N_{\rm sig}$ for $E_{\rm beam}=250$, $500$, and $1500\ {\rm GeV}$.
In Fig.\ \ref{fig:mxvseps}, we plot the contours of constant $N_{\rm
  sig}$ on the $m_X$ vs.\ $\epsilon$ plane. The number of signal is
suppressed for both large and small values of $\epsilon$.  When
$\epsilon$ is too small, the production cross section as well as the
number of the decay inside the decay volume are suppressed.  On the
contrary, with too large $\epsilon$, most of the hidden photons decay
before reaching the decay volume.

Now, we discuss several issues related to the backgrounds.  First, the
muons produced by $e^\pm$-${\rm H_2O}$ scattering may become serious
background.  We estimate the spectrum of the muons produced in the
dump as
\begin{widetext}
\begin{align}
  \frac{d N_{\mu^++\mu^-}}{dp_z}
  = 2 N_e \frac{N_{\rm Avo} X_0}{A}
  \int \frac{d m_{\gamma^*}^2}{\pi}
  \int dE_{\gamma^*}
  \int dE_e 
  \int dt 
  \frac{I_e(E_{\rm beam},E_e,t)}{m_{\gamma^*}^3 E_e}
  \left.
    \frac{d\sigma_0}{dx}
  \right|_{x=E_{\gamma^*}/E_e}
  \frac{d\Gamma(\gamma^*\rightarrow \mu^+\mu^-)}{dp_z},
  \label{Nmu}
\end{align}
\end{widetext}
where $p_z$ is the $z$-component of the momentum of the muon.  In
addition, $d\Gamma(\gamma^*\rightarrow \mu^+\mu^-)/dp_z$ is the
differential decay rate of the ``virtual photon'' with its energy of
$E_{\gamma^*}$ and the invariant mass of $m_{\gamma^*}$.  We found
that $O(10^6)$ muon pairs are produced with the injection of one bunch
train, and that the energy of the produced muons are typically of the
order of $E_{\rm beam}$.  A significant reduction of the flux of these
muons is mandatory.  One possibility of shielding these muons is to
bend them out from the aperture of the vacuum vessel of the decay
volume using magnetic field.  A total field of $B_\perp\sim O(10)\
{\rm Tm}$ is required to bend out $O(100)\ {\rm GeV}$ muons, if the
aperture of the vacuum vessel is $O(1)\ {\rm m}$.  Assuming that the
magnetic field of $O(1)\ {\rm T}$ is available in the shield region,
$L_{\rm sh}$ should be of $O(10)\ {\rm m}$.  The muon shield using the
magnetic field was studied for SHiP experiment \cite{Alekhin:2015byh,
  Anelli:2015pba}, which is a new fixed target experiment proposed in
CERN; it was pointed out that the return fields of a long sequence of
magnets may bend back the muons which have been once bent out.  Thus,
detailed study of the configuration of the magnets for the muon shield
is necessary; we leave the detailed studies of shield and detector
designs for future consideration.  The SHiP collaboration claims that
the muons can be removed using a carefully designed configuration of
magnetic field with $L_{\rm sh}\sim 50\ {\rm m}$
\cite{Anelli:2015pba}.  Here, we use $L_{\rm sh}= 50\ {\rm m}$ in our
study, and assume that muon reduction is possible with magnetic fields
between the dump and the decay volume.

Neutrino- and muon-induced backgrounds may also exist.  Neutrinos and
muons produced in the dump, as well as cosmic rays, may interact
inelastically with the materials surrounding the decay volume,
resulting in the production of long-lived $V^0$ particles, like
$K^0_{\rm L}$.  Their decay products may mimic the charged particles
produced by the decay of the hidden photon.  The amount of $V^0$
particles produced in such a process depends on the experimental
design.

Assuming no background and requiring a few events to claim the
discovery of the signal of hidden photon, the discovery reach is
significantly enlarged by BD$ee$LC, as shown in Fig.\
\ref{fig:mxvseps}; dark photon with its mass of $O(1)\ {\rm GeV}$ or
smaller may be accessed by BD$ee$LC.  Thus, BD$ee$LC will provide a
new possibility to find a signal of hidden particles.

Next, we shortly comment on the beam dump experiment at FCC-$ee$
(which we call ``BD$ee$CC.'')  Adopting the current design of FCC-$ee$
\cite{FCCee}, the number of electrons available for BD$ee$CC is
$O(10^{10})$ per second, which is $3-4$ orders of magnitude smaller
than that for BD$ee$LC.  Even so, BD$ee$CC can enlarge the discovery
reach of hidden particles compared to past experiments.  (See Fig.\
\ref{fig:mxvseps}.)

Finally, we compare BD$ee$ with another possible hidden particle
search in the future, SHiP experiment \cite{Anelli:2015pba}.  The
expected discovery reach of SHiP is also shown in Fig.\
\ref{fig:mxvseps} for the hidden photon model.  We can see that, if
approved, SHiP will also cover the parameter region on which BD$ee$
has a sensitivity.  It should be noted that SHiP is a fixed target
experiment with proton beam, so the fundamental processes producing
hidden particles are different.  If signals of a hidden particle are
discovered, discrimination of various possibilities of hidden
particles may become possible by combining the results of BD$ee$ and
SHiP.

In summary, given the fact that a large number of $e^\pm$ will become
available for beam dump experiment once $e^+e^-$ collider starts its
operation, we propose to install a particle detector behind its dump.
Using the hidden photon model as an example, we have shown that the
beam dump experiment at $e^+e^-$ colliders, BD$ee$, significantly
enlarges the discovery reach of hidden particles. To understand the
potential of BD$ee$, case studies for other hidden particles, like
ALPs and sterile neutrinos, should be performed.  In doing so, the
full capabilities of the machine, such as the use of positrons which
yield annihilation processes, and, in the case of linear colliders,
the use of beam polarization, should be explored.  In addition, the
discovery reach depends on the detail of the configurations of
detectors and shields.  As we have discussed, the muons produced in
the dump are potential serious background and hence careful designs of
detectors and shields are needed.  These issues will be discussed
elsewhere \cite{WorkInProgress}.  BD$ee$ will provide a new
possibility to probe hidden particles, and hence is worth being
considered seriously as an important addition to future $e^+e^-$
facilities.

{\it Acknowledgment}: The authors are grateful to K. Fujii,
K. Nakamura, K. Oda, and T. Suehara for useful discussion.  They also
thank Okinawa Institute of Science and Technology Graduate University
for their hospitality, at which this work was initiated.  This work is
supported by Grant-in-Aid for Scientific research Nos.\ 23104006 (SK),
23104008 (TM), 26400239 (TM), and 23000002 (TT).


\end{document}